%% file: 00_main.tex
\documentclass[journal]{vgtc}                
\ifpdf
  \pdfoutput=1\relax                   
  \usepackage{graphicx}                
  \DeclareGraphicsExtensions{.pdf,.png,.jpg,.jpeg} 
\else
  \usepackage{graphicx}                
  \DeclareGraphicsExtensions{.eps}     
\fi%

\usepackage{microtype}                 
\PassOptionsToPackage{warn}{textcomp}  
\usepackage{textcomp}                  
\usepackage{mathptmx}                  
\usepackage{times}                     
\usepackage{cite}                      
\usepackage{tabu}                      
\usepackage{booktabs}                  

\usepackage[table]{xcolor} 
\usepackage{xcolor}
\usepackage{tabularx}
\usepackage{graphicx}
\usepackage{enumitem}
\usepackage{comment}
\usepackage{float}
\definecolor{cindysalmon}{rgb}{1, 0.55, 0.42}

\usepackage{etoolbox}

\graphicspath{{figures/}{pictures/}{images/}{./}} 

\usepackage{microtype}                 
\PassOptionsToPackage{warn}{textcomp}  
\usepackage{textcomp}                  
\usepackage{mathptmx}                  
\usepackage{times}                     
\usepackage{cite}                      
\usepackage{tabu}                      
\usepackage{booktabs}                  

\onlineid{1100}

\vgtccategory{Research}
\vgtcpapertype{Theoretical and Empirical}

\title{\centering Using Processing Fluency as a Metric of Trust \protect\\ in Scatterplot Visualizations}

\author{Hamza Elhamdadi, Lace Padilla, and Cindy Xiong, \textit{Member, IEEE}}
\authorfooter{
\item
 Hamza Elhamdadi is with UMass Amherst CICS.
 \newline
  Email: helhamdadi@umass.edu
\item
 Lace Padilla is with University of California Merced
 \newline
  Email: lace.padilla@ucmerced.edu
\item
 Cindy Xiong is with UMass Amherst CICS.
 \newline
  Email: cindy.xiong@cs.umass.edu.
}

\shortauthortitle{Xiong \MakeLowercase{\textit{et al.}}: TrustVis Paper Title}

\abstract{Establishing trust with readers is an important first step in visual data communication. But what makes a visualization trustworthy? Psychology and behavioral economics research has found processing fluency (i.e., speed and accuracy of perceiving and processing a stimulus) is central to perceived trust. We examine the association between processing fluency and trust in visualizations through two empirical studies. In Experiment 1, we tested the effect of camouflaging a visualization on processing fluency. Participants estimated the proportion of data values within a specified range for six camouflaged visualizations and one non-camouflaged control; they also reported their perceived difficulty for each of the visualizations. Camouflaged visualizations produced less accurate estimations compared to the control. In Experiment 2, we created a decision task based on trust games adapted from behavioral economics. We asked participants to invest money in two hypothetical companies and report how much they trust each company. One company communicates its strategy with a camouflaged visualization, the other with a controlled visualization. Participants tended to invest less money in the company presenting a camouflaged visualization. Hence, we found support for the hypothesis that processing fluency is key to the perception of trust in visual data communication. 
} 


\CCScatlist{ 
 \CCScat{K.6.1}{Management of Computing and Information Systems}%
{Project and People Management}{Life Cycle};
 \CCScat{K.7.m}{The Computing Profession}{Miscellaneous}{Ethics}
}

 \teaser{
 \centering
 \includegraphics[width=16.5cm]{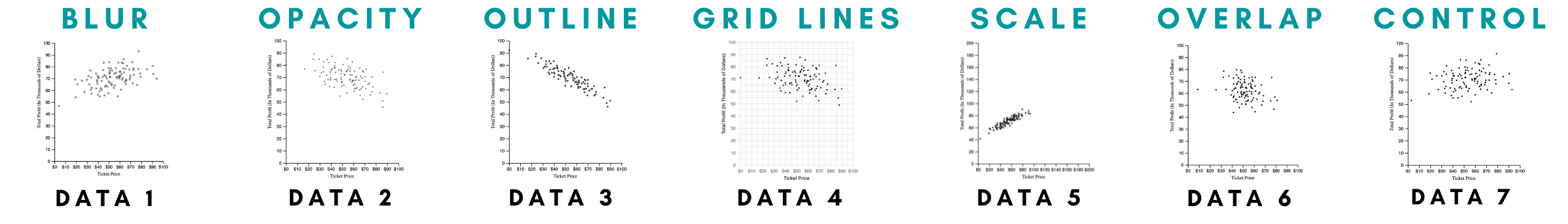}
  \caption{We compared people's performance completing several tasks that include six camouflaged visualizations and one control visualization. The above example shows the camouflaged and control conditions applied to the same dataset.}
  \label{teaser}
  }

\vgtcinsertpkg

\begin{document}


\firstsection{Introduction}

\maketitle


\input{01_Introduction}

\input{03_Overview}

\input{04_Exp1}
\input{05_Exp2}

\input{06_Discussion}


\bibliographystyle{abbrv}

\bibliography{bib}







\end{document}

%% file: 01_Introduction.tex










People are increasingly relying on data visualizations to make decisions, and trust plays a critical role in human-data interactions ~\cite{sacha2016role, dow1998crying}.
Social science research has demonstrated that trust is a dynamic process, in which human readers can learn to trust or distrust conveyed information, and visualizations can earn or lose trustworthiness depending on their design and delivery \cite{boy2015storytelling, xiong2019examining}.
Ideally, we want human readers to engage in ``calibrated trust'' when interacting with data visualizations, which involves \textit{critically evaluating the information, rather than unconditionally dismissing or accepting it} \cite{elhamdadi2022measure}.
At the same time, we want to support visualization creators to design visualizations that elicit calibrated trust.
But how trust impacts human-data interactions has remained mostly under-explored. 
The visualization community does not yet have a systematic understanding of factors that impact trust in visualization design nor a formalized model of how trust is measured and established between humans and data. 

But before investigating the visualization design components that enhance or diminish trust to generate guidelines for promoting trust in visual data communication, we need to identify reliable and valid methods for measuring trust in visual data communication across different contexts and account for individual differences \cite{elhamdadi2022measure}.

Trust has been linked to the halo effect, where a single positive quality of an entity is extrapolated to a generally positive assessment of the entity in other areas \cite{thorndike1920constant, wetzel1981halo}.
For example, people that are perceived as beautiful are also categorized as more trustworthy \cite{kahneman2011thinking}. 
But beauty is a subjective experience that can be difficult to measure \cite{reber2004processing}.
Cognitive science researchers have identified perceptual fluency, or the ease with which a stimulus is perceived, to be positively correlated with trust \cite{oppenheimer2008secret, sohn2017consumer}, mediating the relationship between perceived beauty and trust \cite{oppenheimer2008secret, olszanowski2018mixed, halberstadt2014easy, graf2015dual}.
For example, stocks with easier-to-pronounce names were more likely to be trusted than those with hard-to-pronounce names \cite{alter2006predicting}, when controlled for the size and industry of the companies \cite{alter2006predicting}.
We hypothesize that we can measure the perceived trustworthiness of a visualization by proxy of perceptual fluency, operationalized as the \textit{speed and accuracy} with which one interprets a visualization.

To test our hypothesis, we investigate whether we can influence trust by manipulating the processing fluency of a visualization. 
Specifically, we scope our manipulation to focus on varying the perceptual clarity of visualizations. 
We test several techniques from existing literature and real-world applications to reduce the perceived clarity of visualizations; these techniques include blurring the visual marks, increasing the opacity of visual marks, adding outlines to visual marks, increasing the amount of overlap between visual marks, adding gridlines to the visualization, and manipulating the visualization scale.
We refer to these manipulations as \textbf{\textit{camouflage}}.
We hypothesize that these camouflaged design choices will make visualizations harder to perceive/interpret and decrease processing fluency and trust, which will demonstrate the effect of using processing fluency as a proximate measure of trust.

Existing computer science and visualization research varies in the approach to measuring and defining trust \cite{artz2007survey, mayr2019trust, chita2021can}.
The most prominent measures include a single-item Likert scale \cite{zhou2019effects, xiong2019examining, zhou2019effects} (e.g., ``on a scale from 1 to 7, how much do you trust this visualization?''), with varying range and granularity (e.g., 0 to 100 \cite{kim2017data}, 1 to 7 \cite{jian2000foundations}, etc.).
Other approaches include using substitutions variables as a proxy of trust, such as decisions \cite{zehrung2021vis, xiong2019examining},  perceived credibility and appropriateness \cite{kong2019trust}, or model agreement \cite{yin2019understanding}. 
However, existing work in visualization has rarely tested for the reliability and validity of trust metrics, which are critical to ensuring the accuracy and replicability of results \cite{elhamdadi2022measure}.
We turn to the field of behavioral economics for an additional, more objective way to measure trust beyond a self-report via Likert scale: the Trust Game.

In a typical trust game, two participants are anonymously paired. 
One participant is given an amount of money and told to send some money to the second participant (the amount may be zero). The money that was sent to the second participant gets tripled, and the second participant is told to send some money back to the first participant (the amount can again be zero). The amount sent by the first participant is seen as proportional to the amount of trust they have in the second participant (to return on the investment).

\vspace{2mm}
\noindent \textbf{Contribution:} We synthesize psychology and behavioral economics research with data visualizations by examining how the perceptual fluency of visualizations and trust games from behavioral economics literature might serve as proxies for measuring trust in data visualizations. 
We contribute two empirical studies that compare objective and subjective measures of trust to demonstrate the potential of perceptual fluency and trust games as trust metrics.

In Experiment 1, we test the effectiveness of camouflaged design on decreasing fluency. We measure participants' processing fluency by comparing their performance on a perception task when using visualizations with a camouflaged design to their performance when using a control visualization with no camouflaged design elements.
Our results suggest that some camouflaged designs can decrease processing fluency as compared to the control, although the effect size is small and not statistically significant. We discuss potential follow-ups to further investigate the relationship between camouflage and processing fluency.
In Experiment 2, we examine the relationship between camouflaged design and trust, following a trust game from behavioral economics literature \cite{zurn2017trust}.

%% file: 03_Overview.tex
\section{Design Motivation}
\label{camouflage_types}

We test several techniques from existing literature and real-world applications to manipulate processing fluency by reducing perceived clarity, based on techniques described by Flavell et al. \cite{flavell2020competing}. 
We refer to these manipulations as camouflage.
We created six methods of camouflage: blur, opacity, outline, gridlines, scale, and overlap, based on relevant examples in computer vision, visualization, and psychology literature. As shown in Figure \ref{teaser}, some of these camouflage methods affect only the encoding marks on the visualization, while others (e.g., gridlines and scale) add marks or change the scale of the visualization. 
\newline

\noindent \textbf{Blur:} We referenced a technique from \cite{flavell2020competing}, which adds noise to a visual stimulus to decrease processing fluency. In the visualization community, Gaussian blurring often conveys uncertainty \cite{maceachren2012visual}. Additionally, low resolution in online or print environments can also introduce blur in visualizations. In our experiments, we applied a blur to our visualization stimuli by adding the CSS style attribute ``filter'' with value ``blur(1px)'' to each point in the scatterplot.
\newline
\vspace{-2mm}

\noindent \textbf{Opacity:} Visualization designers often manipulate opacity to increase the visibility of overlapping encoding marks. Designers can also introduce opacity as a visual encoding channel to encode a dimension of data. For example, G{\"u}nther et al. \cite{gunther2014hierarchical} leveraged opacity via a hierarchical approach in linesets to emphasize the importance of certain marks over others, and MacEachren et al., identified opacity as an encoding channel for uncertainty \cite{maceachren2012visual}. In our experiments, we created visualizations where each point had an opacity attribute with a value of 0.5.
\newline
\vspace{-2mm}

\noindent \textbf{Mark outlines:} Designers often add borders to visual elements to emphasize specific data points (e.g., in \cite{schnurer2020empirical}). Some even leverage the thickness of borders as an additional encoding channel (e.g., in \cite{kong2010perceptual}). Adding outlines to encoding marks in visualizations can increase visual contrast between the data and the background \cite{zhang2014visual}, which may impact processing fluency. However, it also adds additional area between encoding marks, increases visual complexity, and may decrease visual differences depending on the thickness of the outlines and the specific colors used \cite{szafir2017modeling}. We created visualizations with outlined marks by giving each point in the scatterplot a ``stroke'' style element with a value of ``grey'' and a ``stroke-width'' style element with a value of 1.
\newline
\vspace{-2mm}

\noindent \textbf{Gridlines:} Visualization designers add gridlines to visualizations as reference points to help people more accurately read data from the visualization. At the same time, however, obtrusive gridlines can clutter the visualization, making the values more difficult to read and decreasing their processing fluency \cite{bartram2011whisper}. We created visualizations with ten evenly-spaced gridlines along each axis with an opacity of 0.7.
\newline
\vspace{-2mm}

\noindent \textbf{Scale:} Re-scaling is a common technique used in visualizations to make differences between data values more perceivable \cite{pandey2015deceptive}, which can likely increase the processing fluency of the visualization. However, the visualization community has debated using this technique due to its potential to deceive readers about effect sizes in data \cite{correll2020truncating, hofman2020visualizing}. A visualization author can either manipulate the scale to reflect the actual effect size at the expense of sacrificing processing fluency or maximize ease of perception (and fluency) at the risk of data misinterpretation. In our experiments, we created scaled visualizations by changing the domain of the x and y axes to [0,200] from the standard [0,100] to manipulate processing fluency.  
\newline
\vspace{-2mm}

\noindent \textbf{Overlap:} Visualization designers also manipulate the amount of overlap to make data more perceivable through jittering.
Readers can struggle to differentiate highly overlapped visual marks, making visualizations with highly overlapped points low in processing fluency \cite{liu2017uncertainty}.
Additionally, readers are more likely to take mental shortcuts when making sense of highly overlapping data, which makes them more prone to making mistakes \cite{padilla2019visualspatial}. 
But overlapping data points occur in real-world visualizations with many data points. 
We created visualizations with overlapping points by manipulating the x and y coordinates of every point on the scatterplot. For data points whose x coordinate was less than 50, we incremented the x coordinate by 20. For data points whose x coordinate was greater than 50, we decremented the x coordinate by 20. We did the same for y coordinates.
\newline
\vspace{-2mm}

\section{Hypothesis and Results Preview}
In this paper, we conduct two experiments. The first experiment explores the effect of various manipulations to processing fluency in data visualizations. The second experiment compares the relationship between processing fluency and trust to examine the viability of using processing fluency as a proxy to measure trust.
\newline
\vspace{-2mm}

\noindent \textbf{Hypothesis No.1:} Camouflaged visualizations will make the visual-perception task (approximating the percentage of data marks within a range) more difficult to complete than non-camouflaged visualizations. 
\newline
\vspace{-2mm}

\noindent \textbf{Hypothesis No.2:} Participants will report using more effort to complete tasks with camouflaged visualizations than control visualizations. 
\newline
\vspace{-2mm}

\noindent \textbf{Hypothesis No.3:} Camouflaged (less fluent) visualizations will be rated as less trustworthy than non-camouflaged (more fluent) visualizations. 
\newline
\vspace{-2mm}

\noindent \textbf{Hypothesis No.4:} In a trust game setting,participants' 
investments (e.g., tickets) will be positively correlated with processing fluency. In other words, non-camouflaged visualizations will receive a higher percentage of tickets than the camouflaged visualizations.
\newline
\vspace{-2mm}

In Experiment 1, we found support for \textbf{Hypothesis No.1.}; however, this support is statistically insignificant. We failed to support \textbf{Hypothesis No.2.}
In Experiment 2, we found support for \textbf{Hypothesis No.3} and \textbf{Hypothesis No.4}.
Generally, these experiments suggest that processing fluency is positively associated with trust in visualized data, making it a potentially viable proxy measurement of trust.



%% file: 04_Exp1.tex
\section{Experiment 1 Comparing Camouflage Types}
Experiment 1 serves as a manipulation check to test whether our camouflage techniques can reduce processing fluency. 
We measure processing fluency by capturing participants' performance accuracy and perceived effort on a perception task. 
Once we validate that our manipulation of fluency works, we can use these less fluent visualizations, in comparison with the more fluent control versions, to investigate the relationship between processing fluency and trust in Experiment 2. 


\subsection{Stimuli}
\label{stimuli}
We created camouflaged visualizations with multiple synthesized, artificial datasets to control for data variability in the visualizations, considering that underlying data values can impact viewer perception \cite{kim2018assessing, xiong2021grouping}. 


We provided each visualization with two labeled axes (seen in Figure \ref{teaser}). We have provided code for the application used for testing at \url{https://osf.io/t7xmh/?view_only=62abec4716464056a5b7fccb7529d6e8}

\subsection{Design}

We used a 7x7 Graeco-Latin Square mixed-subjects design. Each participant saw seven visualizations (six camouflaged, one control). The 49 visualizations are arranged in seven groupings (Figure \ref{teaser} shows an example grouping). The first participant sees the first grouping in a random order, the second participant sees grouping 2 in a random order, and so on. This design structure accounts for order and carryover effects and ensures that not everyone sees the same visualizations in the same order. 

Within every grouping, we present each visualization with a different dataset (generated as described in Section \ref{stimuli}), such that we pair every visualization with every dataset overall. 
We counter-balance with 49 distinct visualizations, each with a single camouflage design-dataset pair. Thus, we minimize the confounding effect of the dataset and visualization design for more generalizable results.

\subsection{Procedure}

The experiment was crowd-sourced via Prolific \cite{palan2018prolific}. Participants were given a series of seven charts following the Graeco-Latin Square (see Figure \ref{teaser} for examples) and asked to complete a perception task and self-report the amount of effort they used to complete the task. 
The perception task asked participants to estimate the proportion of data values within a range in the figure, inspired by questions from visual literacy assessments that yielded the most differentiable results \cite{lee2016vlat}. Specifically, participants answered: What percentage of the tickets are priced between \$x and \$y?.


The variables x and y started at x = 10 and y = 30 for the first chart. We incremented x and y by 10 for each of the other six charts (e.g., x = 20 and y = 40), such that, across the seven groupings, we presented each camouflage technique with questions across a variety of ranges. 
This counterbalancing methodology created variability in the task without impeding task difficulty, enabling us to make more generalizable conclusions.

The self-reported effort question was inspired by the NASA-TLX \cite{cao2009nasa}. To avoid overloading participants with too many questions, we only used the TLX ``effort'' metric, as existing research in the visualization community suggests it to be among the most informative metric in terms of accessing cognitive load \cite{castro2021examining}. We provided a slider below the question in Likert scale, following best-practice recommended in recent psychology methodology work \cite{casper2020selecting}. Specifically, participants were provided the options 1 = ``Very Easy", 2 = ``Moderately Easy", 3 = ``Slightly Easy", 4 = ``Medium", 5 = ``Slightly Hard", 6 = ``Moderately Hard", 7 = ``Very Hard". 
When the participant updated the slider, the text displaying their answer updated with the new slider value (e.g.,``Your Answer: Medium").


When the participant was satisfied with their response to both questions, they clicked the ``Submit" button.
We then collected the participants' demographic information, including their age, gender, distance to monitor, and level of education. 

\subsection{Participants}
\label{exp1_participants}

Based on pilot data collected from 50 participants, we conducted a power analysis that suggested that a target sample of 606 participants would yield 70\% power to detect an overall difference in estimation error (actual percentage of points within range - estimated percentage) between the six different manipulations of camouflage at an alpha level of 0.05, assuming an effect size of approximately 0.13.

We collected data from 612 participants on Prolific \cite{palan2018prolific} filtering for people who are fluent in English, did not complete any of our pilot studies, use only a desktop, and passed the attention check.

\begin{figure}
    \centering
    \includegraphics[width=\linewidth]{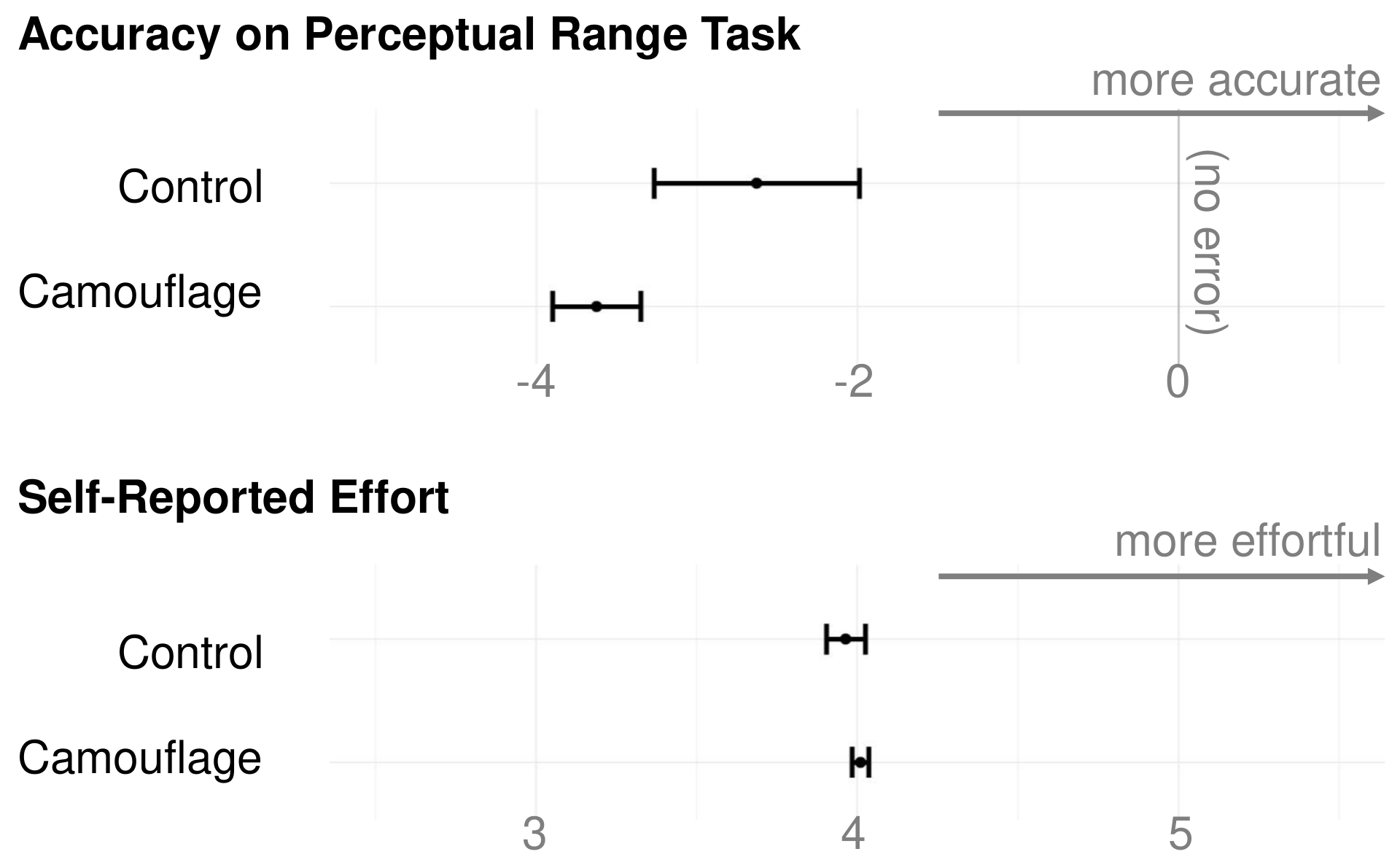}
    \caption{We plot the participants' accuracy on the perceptual data-range estimation task and their self-reported effort. Participants performed significantly worse on the perceptual task when viewing camouflaged visualizations. However, we found no significant difference between reported effort.}
    \label{exp1_results}
\end{figure}

\subsection{Perception Task Results}
We used the lmer function in R \cite{bates2005fitting} to construct a mixed-effect linear model to fit participants' estimation error on the perception task (correct answer - participant response) and the type of visualization shown to the participants (camouflage or control).
We also added the grouping used, age, gender, education level, contrast test response, and distance from monitor as fixed-effect covariate predictors.
We used a random intercept term accounting for individual differences and question index as random effects.

\begin{figure*}
    \centering
    \includegraphics[width=\linewidth]{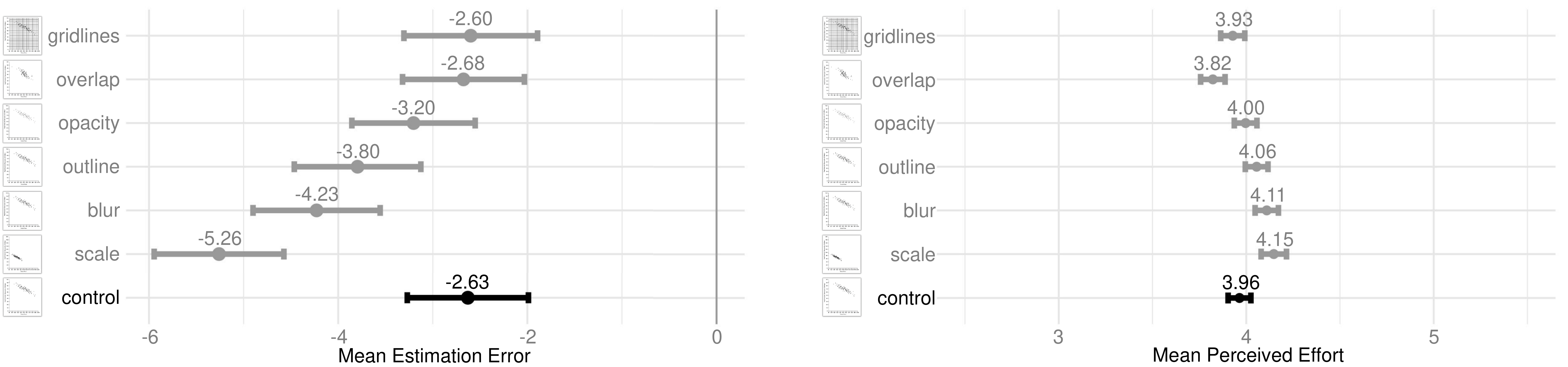}
    \caption{We plotted the average estimation error (left) and perceived effort (right) from Experiment 1 with error bars showing standard error.}
    \label{exp1_results_details_camo}
\end{figure*}

As shown in Figure \ref{exp1_results}, we observed insufficient evidence to support a significant difference between perceptual task accuracy depending on whether the visualization is camouflaged or controlled ($\chi^{2}$ = 2.78,  Est = 1.00, SE = 0.60,  p = 0.096). 
However, this lack of significance may have been due to our treating the different camouflage types as part of the same category when comparing the results to those of the control. Some camouflage types (e.g., scale, blur, outline) produced a very large estimation error compared to the control (see Figure \ref{exp1_results_details_camo}). These differences may have been significant if we compared them separately to control.

We observed no significant effect of the grouping used ($\chi^{2}$ = 0.0091, p = 0.92), age ($\chi^{2}$ = 0.12, p = 0.73), gender ($\chi^{2}$ = 2.51, p = 0.47), education level ($\chi^{2}$ = 0.0060, p = 0.94), response to the contrast test ($\chi^{2}$ = 0.0059, p = 0.94), nor distance from monitor ($\chi^{2}$ = 2.86, p = 0.091).

We compared participants' performance on the perception task with each camouflage technique for sanity. The current study was not powered to detect a significant difference between the camouflage conditions, but, as shown in Figure \ref{exp1_results_details_camo}, participants were less accurate in the perception task when viewing most camouflaged visualizations. Their performance with the gridlines and the overlapping conditions seems comparable to the control. We discuss future opportunities to investigate the effect of different camouflaging techniques on processing fluency in Section \ref{limitation}.



\subsection{Subjective Effort Results}

We also used the lmer function in R \cite{bates2005fitting} to construct a mixed-effect linear model to fit participants' subjective ratings of effort used to complete the estimation task and the type of visualization shown to the participants (camouflage or control).
We added the grouping used, age, gender, education level, contrast test response, and distance from monitor as fixed-effect covariate predictors.
We used a random intercept term accounting for the random effects of individual differences and question index.

From this model, we observed no significant effect on perceived effort depending on whether the visualization is camouflaged or controlled ($\chi^{2}$ = 1.62,  p = 0.20), as shown in Figure \ref{exp1_results}. The more detailed comparison between different camouflage techniques supports this observation (see the right panel of Figure \ref{exp1_results_details_camo}).

We also observed no significant effect of the grouping used ($\chi^{2}$ = 0.96, p = 0.33), age ($\chi^{2}$ = 0.046, p = 0.83), gender ($\chi^{2}$ = 20.78, p < 0.01), education level ($\chi^{2}$ = 1.32, p = 0.25), response to the contrast test ($\chi^{2}$ = 0.065, p = 0.80), nor distance from monitor ($\chi^{2}$ = 1.26, p = 0.26).

\subsection{Discussion}

The results of this experiment indicate that people produce marginally less error (statistically non-significant) when performing the data-range estimation task on non-camouflaged visualizations (the control) than camouflaged visualizations. Considering that processing fluency is primarily defined by the ease of perception \cite{flavell2020competing, oppenheimer2008secret}, this performance suggests that visualization clarity, or lack thereof, impacts processing fluency.

However, people subjectively report similar perceived difficulty completing the data-range estimation task with non-camouflaged and camouflaged visualizations. There are several possible explanations. Existing work has demonstrated that self-reports are not always reliable, considering the cognitive biases people exhibit when making judgments about their abilities \cite{brutus2013self}. Additionally, we conducted our power analysis based on the potential to detect a difference between the control and camouflaged in the perceptual task, rather than the perceived effort. 


Furthermore, although we did not observe any effect of demographic variables such as age, education level, gender, and distance from monitor, individual differences may exist, and we discuss in Section \ref{limitation} future opportunities to further explore the area. 


%% file: 05_Exp2.tex


\begin{figure}
    \centering
    \includegraphics[trim={0.5cm 17.95cm 25.5cm 2.5cm},clip,width=\linewidth]{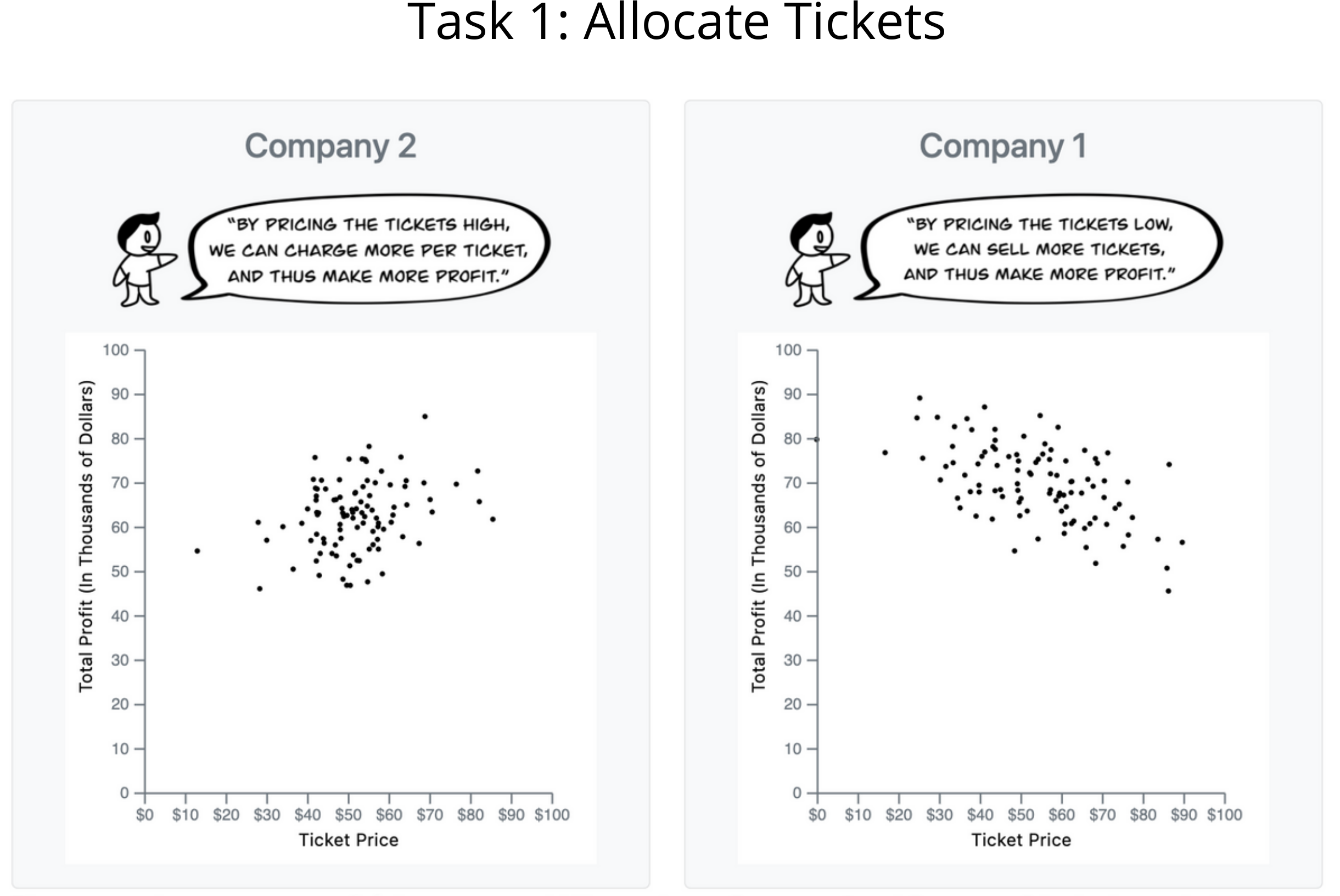}
    \caption{Participants engaged with a hypothetical scenario where they must sell tickets to an event. They divided their tickets between two third-party consulting companies that offered different strategies, shown via camouflaged or controlled visualizations. They then rated the trustworthiness of the companies based on the visualization on a 7-point scale from ``Very Untrustworthy'' to ``Very Trustworthy.''}
    \label{exp2_example}
\end{figure}

\section{Experiment 2 Camouflage and Trustworthiness}
\label{investmentExp}
Experiment 2 investigates the relationship between processing fluency and trust. We operationalize processing fluency as the amount of camouflage, using the camouflaged visualizations (low fluency) in comparison to the control (high fluency) from Experiment 1. We capture two measures of trust: an objective measure where participants make hypothetical investment decisions via a modified trust-game setting (based on \cite{zurn2017trust}), and a subjective measure of trust on a 7-point Likert scale (see Figure \ref{exp2_example}).

\subsection{Participants}

Based on pilot data collected from 50 participants, we conducted a power analysis that suggests that a target sample of 620 participants would yield 70\% power to detect an overall difference between the trust ratings for the camouflaged versus the default visualization at an alpha level of 0.05, assuming a small effect size of approximately 0.10. 
We collected data from 624 participants on Prolific \cite{palan2018prolific}, filtering for people who live in the United States, are fluent in English, did not complete our pilot studies, use only desktops, and passed the attention check (same as described in Section \ref{exp1_participants}).

\subsection{Stimuli and Design }
We used the same visualizations from Experiment 1 as stimuli in this experiment (see Figure \ref{teaser}). We additionally created drawings representing the company recommendations using a digital drawing application.

This experiment used a within-subjects design to compare the effect of the processing fluency on investment decisions and subjectively rated trust. 
Each participant views one control visualization and one camouflaged visualization.
We counterbalanced the applied camouflage technique such that we show each of the six techniques to a subset of participants. 
We also counterbalanced the datasets visualized, the location on the screen the camouflaged visualization appears (left or right), and whether the visualization is associated with the label Company 1 or Company 2.

\begin{figure*} [t]
    \centering
    \includegraphics[trim={0 0.3cm 0 0.1cm},clip,width=\linewidth]{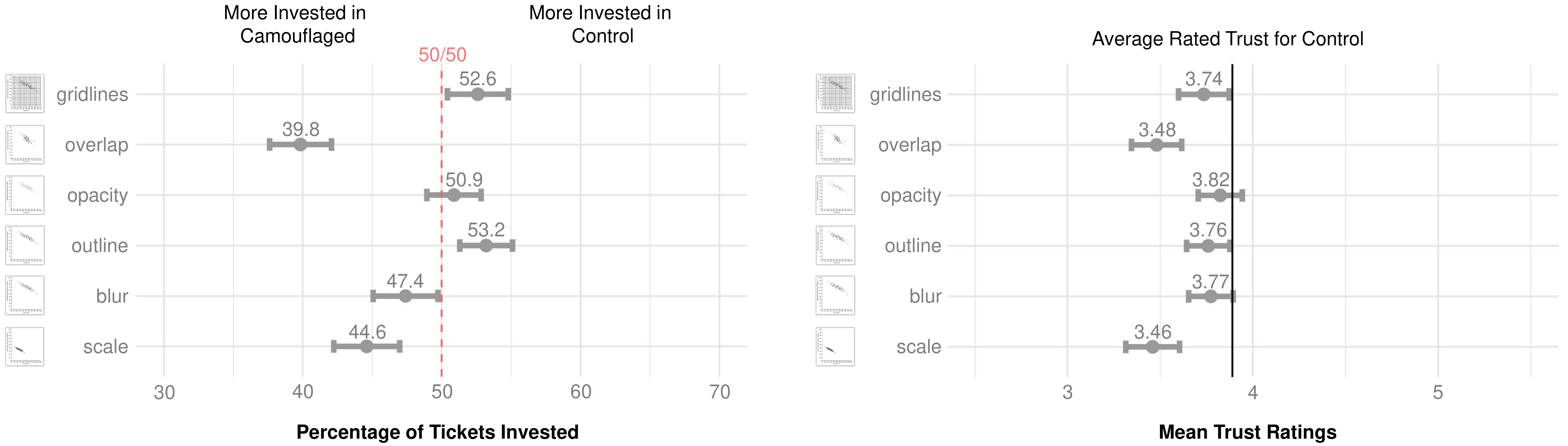}
    \caption{We plotted the average percentage of tickets invested (left) and trust rating (right) with error bars showing standard error from Experiment 2. 
    }
    \label{exp2_results_details_camo}
\end{figure*}

\subsection{Procedure}

Participants first read the consent form and entered their prolific ID, after which they completed an attention check (same as that in Experiment 1) to ensure they were providing attentive responses. If the participant did not answer the attention check correctly, they were disqualified from the study and not permitted to proceed.

After the attention check, we provided the participant with a succinct paragraph of instruction for the study describing the investment decision task.
After reading the instructions, participants could click on ``close'' to proceed to the investment task.
During the investment task, participants viewed two visualizations depicting ticket-selling strategies from two companies.
One company showed a positive correlation between ticket price and total profit, suggesting that the higher you price the tickets, the more profit you can make per ticket and, therefore, in total. The other company showed a negative correlation between ticket price and total profit, suggesting that you can sell more tickets by pricing tickets lower, which will net you higher total profit. 

The application prompted participants to invest a proportion of their tickets with each company. The web application automatically ensured that the percentages for the two companies add to 100.
Once they submitted their investment decision, the application prompted them to rate the perceived trustworthiness of each company via a 7-point Likert scale (1 = ``A Little Trustworthy"; 7 = ``To A Large Extent Trustworthy")

After the study, the participant answered a short four-question survey to determine their graph literacy \cite{okan2019using}. Finally, we collected the participant's demographics using the same demographic questionnaire as those in Experiment 1.

\begin{figure*}[t!]
\centering
\includegraphics[trim={0 3cm 0 25cm},clip,width=\linewidth]{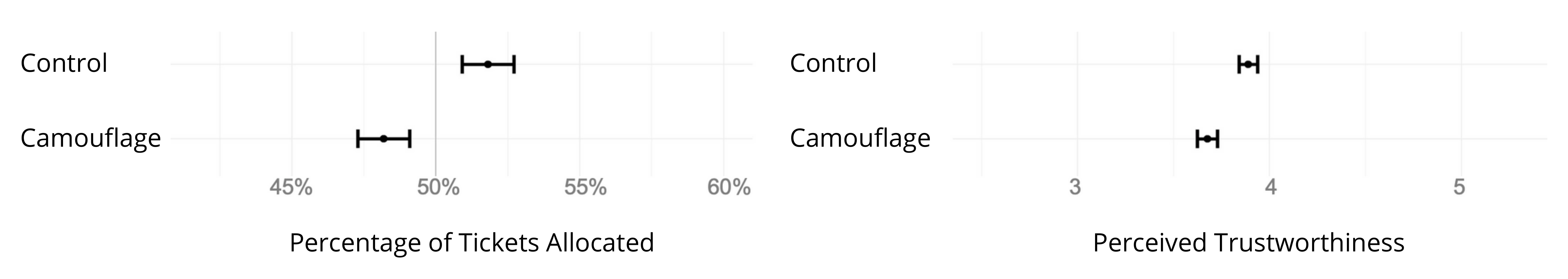}
\caption{Participants invested more tickets in the non-camouflaged visualizations. They also rated the non-camouflaged visualizations as more trustworthy. }
\label{exp2}
\end{figure*}

\subsection{Investment Task Results}

We compared the number of tickets participants reported to invest in the two companies. 
One company showed its data via the control visualization, and the other via the camouflaged visualization.
There is a significant difference between the percentage of tickets invested in the company with the camouflaged visualization and the company with the control visualization (t(1246) = -2.85, p = 0.0044).
On average, participants report wanting to invest more tickets (M = 51.81\% of tickets, SD = 22.45) to the company with the control visualization than to the company with the camouflaged visualization (M = 48.18\%, SD = 22.45), see Figure \ref{exp2}.

Similar to Experiment 1, we compared the amount of tickets participants invested in each camouflage design for sanity. Although the current study is not powered to detect a significant difference between the camouflage conditions, participants seem to invest less in the camouflaged techniques overall, as shown in Figure \ref{exp2_results_details_camo}. Like in Experiment 1, their investment in the gridlines condition seems comparable to the control. Additionally, investment in the outline and opacity conditions also appears similar to the control. We discuss future opportunities to investigate the effect of different camouflaging techniques on investment decisions in trust games in Sections \ref{limitation}.

\subsection{Perceived Trustworthiness Results}

We compared the trust ratings that participants reported for each of the two companies. 
One company showed its data via the control visualization, and the other via the camouflaged visualization. 
We observed a significant difference between the trust ratings for the two companies (t(1238.1) = -2.98, p = 0.0030). 
On average, participants rated the company with the control visualization 3.89 (SD = 1.21, i.e., between ``Neither Trustworthy nor Untrustworthy" and ``Moderately Trustworthy" closer to the latter) and the company with the camouflaged visualization 3.68 (SD = 1.31, i.e., between ``Neither Trustworthy nor Untrustworthy" and ``Moderately Trustworthy" closer to the latter), as shown in Figure \ref{exp2}.

We also looked at the subjective trust ratings of each camouflage condition separately for sanity, as shown in Figure \ref{exp2_results_details_camo}. Overall, we can see that participants seem to trust the control more than any camouflage condition. 
Additionally, the trust ratings seem to generally correlate with the investment decision, which we more closely examine in the next section. 

\subsection{Regression Models Predicting Trust with Investment}

We constructed a linear regression model to predict trust ratings using the percentage of tickets the participants invested, the visualization type (camouflaged or control), and the interaction between them. 
We also tested the covariance of literacy scores, camouflage technique, age, gender, distance from monitor, and contrast test response, to control for their effects.
We added the participant as a random effect because the same participant provided ratings and ticket percentages for control and camouflage.

Overall, we found the percentage of tickets invested to be a significant predictor of trust ratings (Est = 0.0269, SE = 0.0020, $\chi^{2}$ = 459.73, p $<$ 0.001), such that the higher the percentage of tickets invested, the higher trust rating. 

We also found the visualization type (camouflage or default) to be a significant predictor of trust ratings (Est = 0.17, SE = 0.17, $\chi^{2}$ = 4.50, p = 0.034), such that participants rated camouflage visualizations (Mean = 3.68, SE = 0.053) to be less trustworthy than the control visualization (Mean = 3.89, SE = 0.049)

We found no significant interaction between the percentage of tickets invested and visualization type ($\chi^{2}$ = 0.11, p = 0.74). We also found no significant effect of literacy score ($\chi^{2}$ = 0.0005, p = 0.98), camouflage type ($\chi^{2}$ = 2.32, p = 0.80), age ($\chi^{2}$ = 1.03, p = 0.31), gender ($\chi^{2}$ = 3.05, p = 0.69), distance from monitor ($\chi^{2}$ = 0.21, p = 0.65), contrast test response ($\chi^{2}$ = 1.48, p = 0.22), nor screen dimensions ($\chi^{2}$ = 0.0087, p = 0.93). 




\subsection{Discussion}

Generally, people tend to invest more tickets with a company that pitches its strategy via the control (non-camouflaged) visualization. They also report trusting the company more. Together with Experiment 1, which showed that the control visualization is processed more fluently, these results suggest visualization clarity increases processing fluency and, thus, trust. Therefore, measuring processing fluency may, in fact, be a functional proxy for measuring trust in visualizations.





%% file: 06_Discussion.tex
 \section{General Discussion}
\label{discussion}

Since trust plays a significant role in how people perceive scientific information and make critical decisions with data~\cite{sacha2016role, dow1998crying}, examining trust in data communication is becoming a pressing issue for all scientific communities. 
As news outlets increasingly leverage data visualization to communicate scientific findings to the general public \cite{boy2015storytelling}, understanding how the visual communication of data can influence trust in science is an essential step in identifying methods to help people reason about the trustworthiness of findings.

We found evidence showing that camouflaging a visualization is associated with decreased processing fluency and, thus, decreased perceived trust.  This result indicates that techniques from prior work on manipulating fluency through camouflage can be applied to data visualizations to impact trust~\cite{flavell2020competing, oppenheimer2008secret}. 
This connection between processing fluency and trust in visualizations provides a practical method for measuring trust in visualizations: visualizations that have higher processing fluency are more likely to be trusted.

Additionally, we see emerging interest in the visualization community to conduct hypothesis-driven empirical studies investigating human visual perception and intelligence for data visualizations.
The growing body of interdisciplinary work between human perception and data visualization has significantly advanced our understanding of underlying mechanisms of how humans perceive visualized data, contributing concrete design guidelines towards more effective visualizations \cite{Cleveland1984GraphicalMethods, elliott2020design, franconeri2021science}.
The present work suggests that principles derived from human perception studies in data visualizations are also guidelines to increase trust in visual data communication.
Techniques that make visualizations more perceivable and more easily understood increase the processing fluency of a visualization, which is associated with increased trust in the underlying data.
For example, existing work has shown that redundant encoding, the practice of encoding visual marks via multiple channels such as color and shape, can enhance perception. People can more quickly segment objects within a dense display when they are redundantly encoded \cite{nothelfer2017redundant}.
Thus, redundant coding is likely associated with processing fluency, and leveraging this technique in visualization design can potentially increase trust. 
Hence, we strongly urge visualization designers to leverage guidelines produced by visual perception studies to promote trustworthy data communication.

At the same time, this work is one first step towards establishing a more objective metric of trust beyond current practices in data visualization research. Future work should further perfect this method and explore the practical and ethical implications of crafting design guidelines that an enhance trust in visual data communication.



\section{Limitations and Future Directions}
\label{limitation}
We discuss several limitations in our study that provide promising future research directions. We did not cover all possible techniques to create camouflaged visualizations or other aspects of visualization that decrease processing fluency. A rich literature at the intersection of psychology and visualization demonstrates how some designs better facilitate visualization perception \cite{borkin2013what, franconeri2021science, hullman2011benefitting}, and further research should explore the connection between perceptual design decisions and trust in visualization.

More specifically, future research can work towards concrete metrics that measure the fluency of a visualization.
We used only one perception task and one subjective effort task to evaluate visualization fluency. Comprehensive visual analytics/literacy tasks exist to measure a person's ability to interpret visualizations \cite{lee2019correlation, amar2004knowledge}. 
Future work should develop systematic and comprehensive measures of perceptual fluency in visualizations.
These measures will inform actionable design guidelines and allow visualizations authors to revise their designs toward more trustworthy communication. 

We also did not power our experiments to detect differences in how each camouflage design affects trust in visualizations. We observed trends of some camouflaging techniques introducing more dis-fluency than others. Future studies can run additional controlled studies to compare the camouflage designs we tested and generate rankings on how each impacts processing fluency and trust. 
We also did not balance the camouflage manipulations to be equally salient between the conditions. For example, the gridline camouflage was a weaker camouflage compared to the scale camouflage.
Future work can further tweak the camouflage manipulations to ensure more fair comparison of their processing fluency. 

Additionally, although we considered the demographic information we collected from our participants (e.g., literacy, age, gender) in our regression models, we did not design our experiment to identify how individual demographics might impact trust perception in visualization. We suspect individual differences may play a role in how people perceive visualizations as trustworthy, which future work should continue to explore.

Our research focused on the impact of processing fluency on trust in visualization, particularly for scatterplots. We can potentially intuit that lower processing fluency would decrease trust in other types of visualizations, such as line charts, interactive visualizations, bar charts, and dashboards, but further research should empirically test such hypotheses \cite{kosara2016empire}.